\documentclass[aps,prl,twocolumn,groupedaddress]{revtex4-1}

\usepackage{graphicx,amsmath}

\begin{document}


\title{Spin-spiral inhomogeneity as the origin of ferroelectricity in orthorhombic manganites}


\author{I. V. Solovyev}
\email{SOLOVYEV.Igor@nims.go.jp}
\affiliation{
National Institute for Materials Science, 1-2-1 Sengen, Tsukuba,
Ibaraki 305-0047, Japan}


\date{\today}

\begin{abstract}
We argue that the homogeneous distribution of spins in the
spin-spiral state of orthorhombic manganites is deformed
by the relativistic spin-orbit interaction.
The thus induced spin-spiral inhomogeneity
gives rise to the ferroelectric response of
the
purely electronic origin in even-periodic magnetic structures.
The mechanism is generic and explains the appearance of
ferroelectricity
in
the twofold periodic structure of HoMnO$_3$ as well as the
fourfold periodic structure of TbMnO$_3$.
Nevertheless, odd-periodic magnetic structures
preserve the inversion symmetry and
thus are not ferroelectric.
Our analysis is based on the low-energy model derived
from the first-principles electronic structure calculations and
the Berry-phase formalism for the
electronic polarization.
\end{abstract}


\maketitle

  The possibility of switching the electric polarization ${\bf P}$ by means of
magnetic field and the spin magnetization by means of electric field has attracted
enormous attention to multiferroic materials because of their potential
applicability in new electronic devises as well as
the fundamental interest to the problem
of coupling between magnetic and electronic degrees of freedom~\cite{MF_review}.
The orthorhombic rare-earth manganites $R$MnO$_3$ (the space group $Pbnm$)
are
regarded as the key materials for understanding details of such coupling.
After discovering the switching phenomena in TbMnO$_3$~\cite{KimuraNature},
an unprecedented number of investigations has been carried out in order to
clarify magnetic, structural, and ferroelectric properties of various
$R$MnO$_3$ compounds (e.g., Refs.~\cite{Arima,ExpSpinSpiral,nonmagneticRE}).

  All of them are
\textit{improper} ferroelectrics, where the inversion symmetry
is broken due to some complex magnetic ordering~\cite{MF_review}.
Since
the Mn-atoms in the $Pbnm$ structure
occupy the inversion centers, the latter can be destroyed
only if the magnetic unit is larger than the crystallographic one.
There are two types
of compounds with different periodicity,
which are typically
regarded as representatives of two main theories
of ferroelectricity in $R$MnO$_3$. The first one is (nearly)
fourfold periodic TbMnO$_3$, where the ferroelectric activity
is believed to be due to the spin-orbit interaction (SOI)
related spiral spin alignment~\cite{SStheories}.
Another one is twofold periodic HoMnO$_3$,
where ${\bf P}$ is typically ascribed to the (independent on the SOI)
magnetostriction effect in the noncentrosymmetric E-type
antiferromagnetic (AFM) structure~\cite{SergienkoPRL}.

  Nevertheless, many questions remain.
(1) There is no unique theory of multiferroicity in $R$MnO$_3$.
Do HoMnO$_3$ and TbMnO$_3$
really behave as completely different systems, where
the ferroelectricity is caused by different microscopic mechanisms?
Is it possible to unify these two cases?
(2) When the temperature decreases,
many spin-spiral manganites, including TbMnO$_3$, exhibit a lock-in transition into a
fourfold periodic
commensurate
phase~\cite{Arima}. What is the origin of these commensurability?
Is it different from the
twofold periodicity in HoMnO$_3$?
(3) The experimental magnetic structure, derived for TbMnO$_3$ and some related compounds
by assuming the ``spin-spiral'' model, rises many questions: even at low temperature,
in order to fit the experimental data, one had to rely on the
elliptical deformation of the spin spiral~\cite{ExpSpinSpiral}.
However, in the elliptical distribution, the magnetic moments at certain Mn-sites
are substantially reduced (up to about $3 \mu_B$~\cite{ExpSpinSpiral}),
which clearly contradicts to
the Hund's rule physics. Thus, the spin-spiral model is probably incomplete.
If so, what is the true magnetic ground state of $R$MnO$_3$ and how is it related
to the ferroelectric activity of these systems?

  In this work we will rationalize some of these questions. We will argue that
the ground state of $R$MnO$_3$ is \textit{not} the spin spiral. Instead, we will
introduce the concept of the \textit{inhomogeneous} spin-spiral, where the
inhomogeneity is driven by the relativistic SOI and is actually responsible for the
ferroelectric activity in $R$MnO$_3$.
We will argue that this
concept is applicable to all even-periodic systems, including TbMnO$_3$ and HoMnO$_3$.

\begin{figure}
\begin{center}
\includegraphics[height=3.5cm]{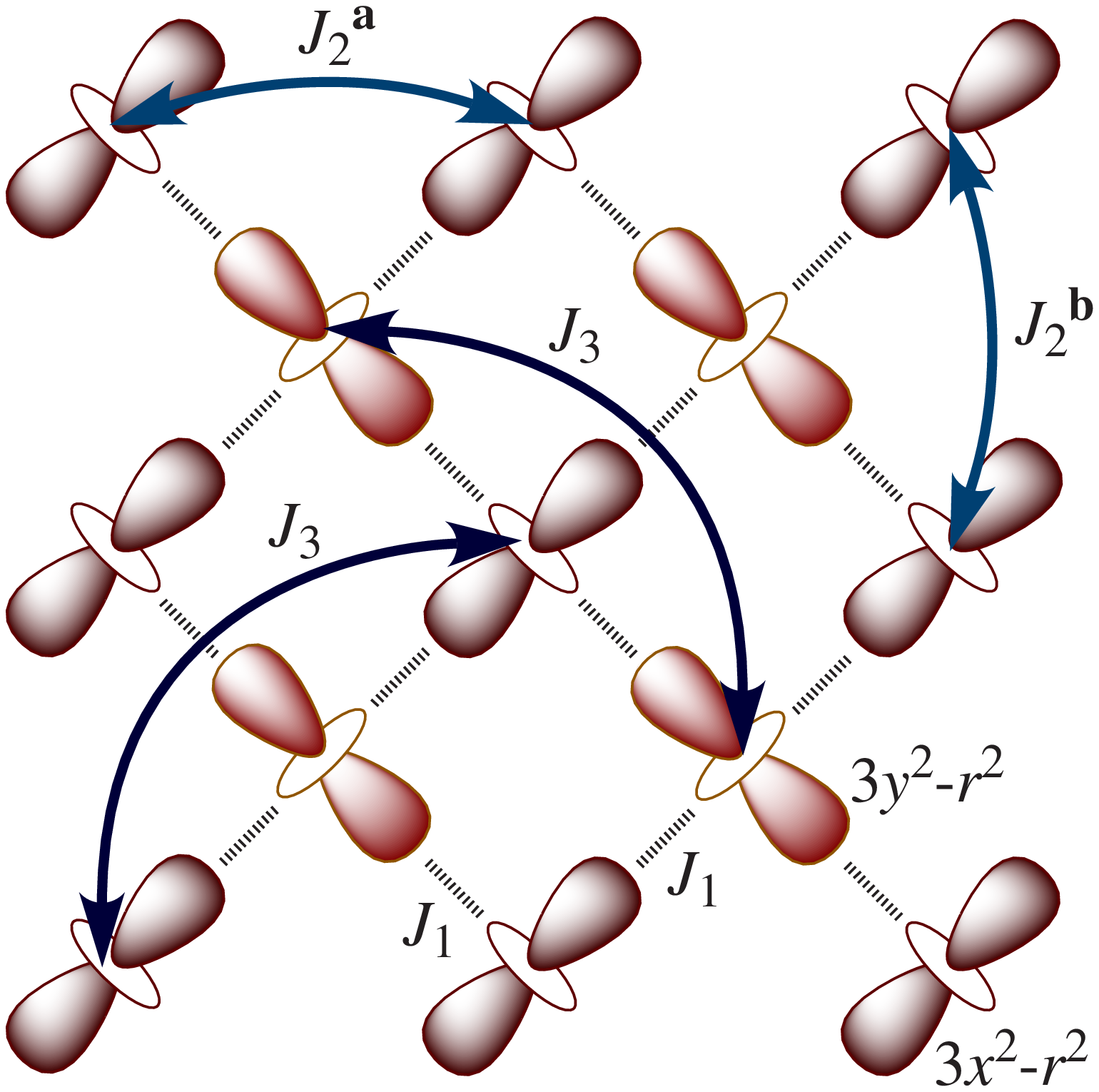}
\includegraphics[height=3.5cm]{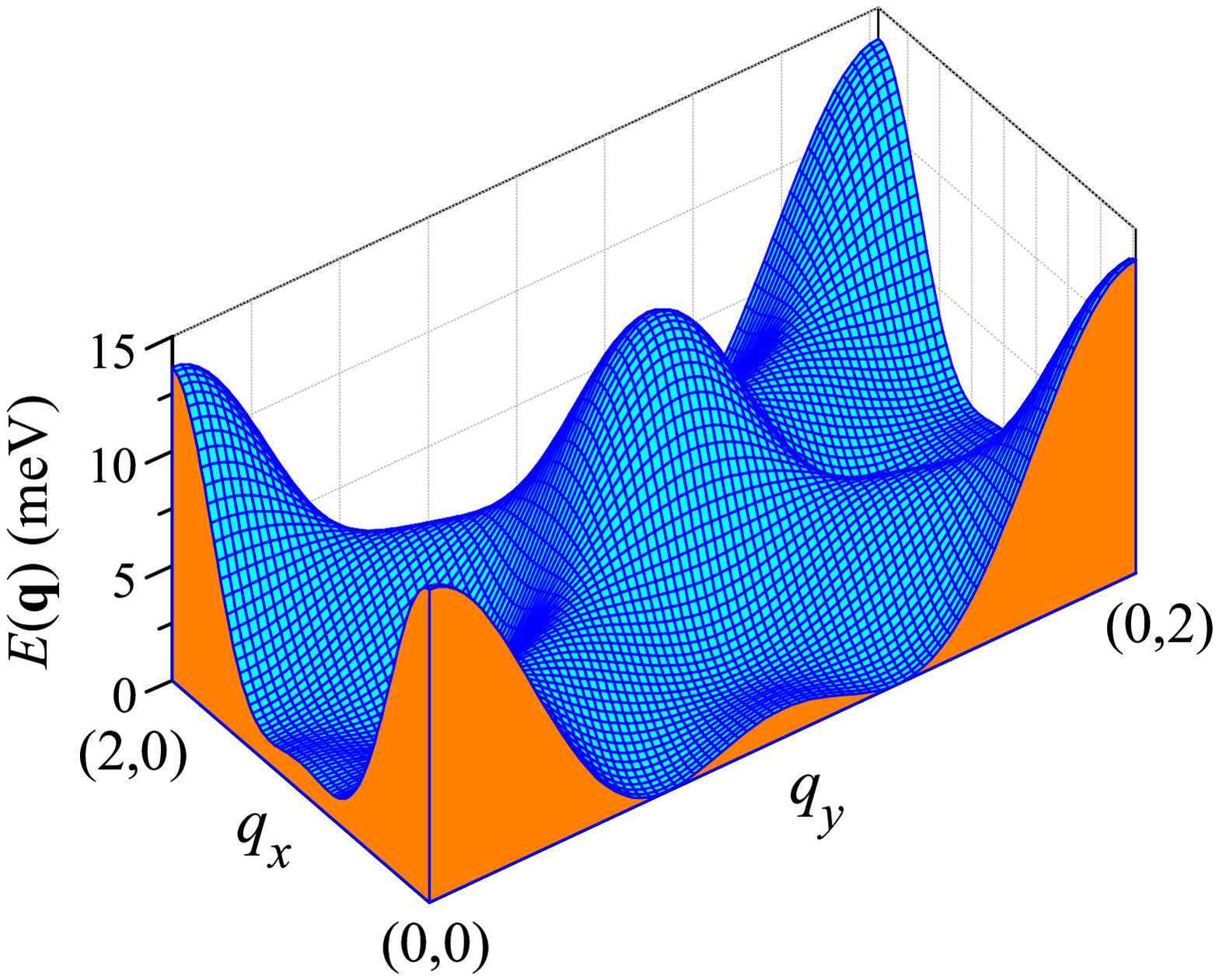}
\end{center}
\caption{\label{fig.noSOI} (Color online) Left panel: schematic view on the
orbital ordering and magnetic interactions in the ${\bf ab}$-plane
of $R$MnO$_3$. Typical values of magnetic interactions
for TbMnO$_3$ are $J_1$$= -$$3.7$, $J_2^{\bf a}$$= -$$0.2$, $J_2^{\bf b}$$= -$$1.2$,
and $J_3$$= -$$2.3$ meV~\protect\cite{JPSJ,remark2}. Rigt panel:
dependence of the total energy
(per one Mn atom)
on the uniform spin-spiral
vector ${\bf q} = (q_x,q_y,1)$
obtained in the HF calculations for TbMnO$_3$ without
SOI.}
\end{figure}

  The magnetic properties of manganites can be linked to the
electronic structure of the Mn$3d$-bands located near the Fermi level.
Thus, these bands, after the transformation to the real space,
can serve as the Wannier-basis for an effective low-energy model.
For $R$MnO$_3$ this
basis includes three $t_{2g}$ and two $e_g$ orbitals
per spin for each of the four Mn-sites in the unit cell.
The model Hamiltonian is taken in the Hubbard form, where all the
parameters,
such as the crystal field, SOI, transfer
integrals, and the effective Coulomb interactions are calculated
rigorously by starting from the local-density approximation (LDA) and
using the experimental crystal structure~\cite{ExpStructure}.
The rare-earth $4f$ states are treated as the core, which does not
contribute to the low-energy properties.
The details
of the computational procedure can be found in the review article~\cite{review2008}.
The results of such calculations for the whole series of the $R$MnO$_3$ compounds without SOI
have been reported in Ref.~\cite{JPSJ}.
Here, we only emphasize two
points, which are important for the magnetic inversion
symmetry breaking:
(1) the Jahn-Teller distortion (JTD) gives rise to the large
($\sim$1.5 eV) crystal-field splitting between $e_g$ levels, which manifests
itself in the orbital ordering (Fig.~\ref{fig.noSOI});
(2) The on-site Coulomb repulsion $U$ is not particularly strong
($\sim$2.2 eV) due to the very efficient screening by the O$2p$ band~\cite{JPSJ},
which is important for the ``right'' balance between
nearest-neighbor (NN) and some longer-range (LR) magnetic interactions,
whose form is controlled by the JTD~\cite{JPSJ,remark2}.

  Besides JTD, the
NN interactions in the
${\bf ab}$-plane depend on the buckling of the Mn-O-Mn bonds.
In the least distorted LaMnO$_3$, $J_1$ is ferromagnetic (FM)~\cite{PRL96,JPSJ}.
However, as the hybridization between $t_{2g}$ and $e_g$ states increases
in the more distorted compounds, $J_1$ can become AFM~\cite{JPSJ}.
Since the on-site Coulomb repulsion is not particularly strong, besides conventional
superexchange, there are other interactions, which appear
in the higher orders of $1/U$, connect more remote sites,
and compete with $J_1$. Among them, the third-neighbor
interaction $J_3$, operating via unoccupied $e_g$ states of
intermediate Mn-sites, is the strongest one. The second-neighbor interactions
$J_2^{\bf a}$ and $J_2^{\bf b}$ are weaker. Nevertheless,
what important is the
anisotropy $| J_2^{\bf b} | \gg | J_2^{\bf a} |$, which predetermines
the direction of propagation (${\bf b}$) of the spin-spiral and the AFM E-state~\cite{JPSJ}.
The NN interactions between adjacent ${\bf ab}$-planes
are strongly AFM (about $-$$8$ meV). The numerical values of $J$'s
for
TbMnO$_3$ are listed in Fig.~\ref{fig.noSOI}.
Similar behavior was found for HoMnO$_3$~\cite{JPSJ}.

  Without SOI, the competition of isotropic interactions in the ${\bf ab}$-plane
gives rise to the incommensurate
spin-spiral state (Fig.~\ref{fig.noSOI}), which can be obtained by applying the generalized Bloch theorem~\cite{Sandratskii}
and solving the model in the Hartree-Fock (HF) approximation~\cite{remark3}.
Due to the AFM interactions between the planes,
the spin-spiral vector ${\bf q}$, which specifies
the phase
$\varphi^{\boldsymbol{\tau}}_{\bf R} = {\bf q} \cdot ( \boldsymbol{\tau}$$+$${\bf R}) + \alpha^{\boldsymbol{\tau}}$
of the direction of spin
${\bf e}^{\boldsymbol{\tau}}_{\bf R} =
\left( \cos \varphi^{\boldsymbol{\tau}}_{\bf R},
\sin \varphi^{\boldsymbol{\tau}}_{\bf R}, 0 \right)$
at the Mn-site $\boldsymbol{\tau}$ of the unit cell ${\bf R}$, can be searched in the form
${\bf q}$$=$$(q_x,q_y,1)$, in units of reciprocal lattice
translations. The total-energy minimum corresponds to the
homogeneous ($\alpha^{\boldsymbol{\tau}}$$=$$0$)
propagation along the orthorhombic ${\bf b}$-axis with $q_y$$=$$0.68$ for TbMnO$_3$
(and $q_y$$=$$0.72$ for HoMnO$_3$),
which exceeds the experimental values $q_y$$=$$0.28$ and $0.25$, reported for
the ${\bf bc}$~\cite{KimuraNature} and ${\bf ab}$~\cite{Arima} helix, respectively.
This discrepancy will be resolved later by considering
the relativistic SOI, but
first we turn to the analysis of the electronic polarization,
which is believed to be closely related to the
spin-spiral alignment~\cite{SStheories}.
For these purposes we compute ${\bf P}$
from the HF eigenvectors $| C_{n {\bf k}} \rangle$ in the
Wannier-basis by using the Berry-phase formalism on the discrete grid of
${\bf k}$-points~\cite{PBerry}. Since the Wannier-basis (and the model itself)
was constructed by starting from the LDA bandstructure,
which respects the inversion symmetry, it does not contribute to ${\bf P}$.
The details can be found in Ref.~\cite{BiMnO3}.
First, we confirmed that without SOI,
the homogeneous spin-spiral state does not produce the electronic polarization.
A finite ${\bf P}$ can be indeed obtained
by switching on the relativistic SOI and performing one iteration
by starting from the self-consistent
nonrelativistic HF potential.
As a test example,
let us consider the $q_y$$=$$\frac{1}{3}$ spiral in TbMnO$_3$, which was
intensively discussed in the literature~\cite{TbMnO3LDAU}.
Then, for
the ${\bf bc}$ and ${\bf ab}$ helix structure,
we obtain ${\bf P} || {\bf c} \sim$ 3 $\mu$C/m$^2$ and
${\bf P} || {\bf a} \sim$ 470 $\mu$C/m$^2$, respectively. Thus,
we confirm that our minimal model successfully reproduces
results of the first-principles
LDA$+$$U$ calculations for the spin-spiral state~\cite{TbMnO3LDAU}:
(1) the inequality ${\bf P} || {\bf a} \gg {\bf P} || {\bf c}$, which
holds for the ${\bf ab}$ and ${\bf bc}$ helix structures;
(2) the absolute values of ${\bf P}$, which are not particularly large.

  Being encouraged by this good agreement,
we turn to the central part of our work, where we will show that
(1) the homogeneous spin-spiral state in
$R$MnO$_3$ is \textit{magnetically} unstable as it
tends to deform to some new \textit{inhomogeneous} state under the SOI;
(2) this inhomogeneity gives rise to the new \textit{electronic} contribution to ${\bf P}$.

\begin{figure}
\begin{center}
\includegraphics[width=6.5cm]{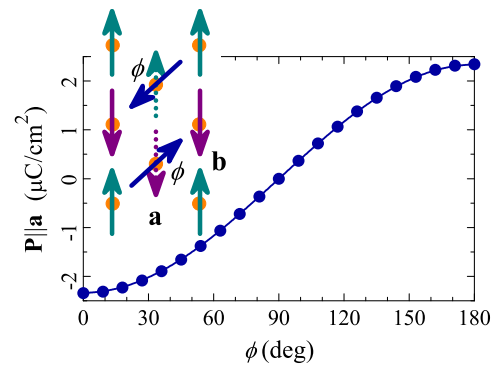}
\end{center}
\caption{\label{fig.2b}(Color online) Nonrelativistic electronic polarization
for the general periodic $q_y$$=$$\frac{1}{2}$ structure of TbMnO$_3$ depending on the angle $\phi$
between spin magnetic moments in two Mn-sublattices. Relative directions of spins
in the ${\bf ab}$ plane are explained in the inset. The numerical value of
$| {\bf P} |$ in the E-state ($\phi$$=$ 0$^\circ$ and 180$^\circ$) of TbMnO$_3$ is
2.3 $\mu$C/cm$^2$ (and 2.7 $\mu$C/cm$^2$ in the
E-state of HoMnO$_3$).}
\end{figure}

  The simplest example, which illustrates how the ferroelectricity
can be induced by the spin-spiral inhomogeneity \textit{without} SOI is
the general $q_y$$=$$\frac{1}{2}$ periodic structure (Fig.~\ref{fig.2b}).
In this case one can realize the collinear (and inhomogeneous in the sense that
it is characterized by two magnetically different Mn-O-Mn bonds)
E-state and the homogeneous spin-spiral
state, depending on the angle
$\phi$
between two Mn-sublattices in the ${\bf ab}$-plane:
$\phi$$=$$0$ and $180^\circ$ give rise to
two E-phase domains with opposite polarization,
while $\phi$$=$$90^\circ$ corresponds to the homogeneous spin-spiral structure
with zero net polarization.
Thus, in order to obtain finite ${\bf P}$,
it is sufficient to ``perturb'' the homogeneous spin-spiral
state in the direction of the inhomogeneous E-state.
For the collinear E-state this mechanism was proposed by Sergienko \textit{et al.}~\cite{SergienkoPRL}.
Then, Picozzi \textit{et al.} considered the behavior of
polarization for the more general noncollinear alignment and on the basis of
first-principles calculations argued that there is a large electronic
contribution to ${\bf P}$~\cite{Picozzi}, which is semiquantitatively reproduced
by our model calculations in Fig.~\ref{fig.2b}.

\begin{figure}
\begin{center}
\includegraphics[width=7.5cm]{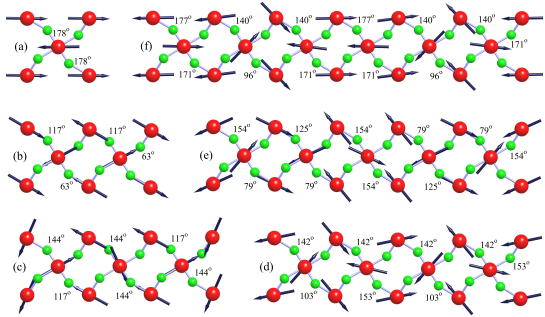} 
\end{center}
\caption{\label{fig.SOGS}(Color online)
Spin patterns in the ${\bf ab}$ plane of TbMnO$_3$ obtained in the
HF calculations with SOI for $q_y$$=$1 (a), $\frac{1}{2}$ (b),
$\frac{1}{3}$ (c), $\frac{1}{4}$ (d), $\frac{1}{5}$ (e), and $\frac{1}{6}$ (f).
The numbers indicate the angles between Mn-spins in different Mn-O-Mn bonds
along the orthorhombic ${\bf b}$ axis.}
\end{figure}

  Now the question is
how to stabilize this
inhomogeneous spin-spiral structure?
One possibility is of course the nonrelativistic exchange striction, which leads to the
off-centrosymmetric atomic displacements and stabilizes the AFM E-state~\cite{Picozzi,SergienkoPRL}.
We will come back to the analysis of relative roles played by different mechanisms
at the end of the paper, but first we want to show that the inhomogeneous magnetic structure,
which gives rise to the finite ferroelectric polarization, can be naturally stabilized by the
relativistic SOI. This mechanism is very generic and takes place in the twofold
periodic HoMnO$_3$ as well as fourfold periodic
TbMnO$_3$.
Since the generalized Bloch theorem is no longer
valid in the relativistic case,
we consider the supercell geometries corresponding to
$q_y$$=$$1/L$ for which
we start from the
homogeneous spin-spiral state, turn on
SOI and further iterate the HF equations until self-consistency. Typically,
this procedure requires
several tens of thousands of iterations, which are accompanied by
the decrease of the total energy. The results
of such calculations
for TbMnO$_3$ are summarized in Fig.~\ref{fig.SOGS}.
The equilibrium magnetic structures are characterized by an alternation of the angles
formed by spins in different Mn-O-Mn bonds, which indicates the inhomogeneity.
The obtained spin patterns reflect the
competition of many interactions in the system: the isotropic
interactions tend to form the uniform spin spiral, while relativistic SOI, which
couples spin magnetic moments to the lattice, tends to restore the commensurability in the system.
The magnetic moments
lie mainly
in the ${\bf ab}$-plane (apart from small canting in the ${\bf c}$-direction)~\cite{remark1}.
The $q_y$$=$$1$ structure is mainly formed by the NN interactions and the
magnetic anisotropy~\cite{PRL96}, while the LR interactions are not affective
due to the periodicity constraint. For the $q_y$$=$$\frac{1}{2}$ structure, the spins are aligned
parallel to the longest Mn-O bonds and minimize the single-ion anisotropy energy.
Moreover, the magnetic coupling along the ${\bf b}$-axis minimizes the energy of LR interactions.
For the large-$L$ structures, the situation is even more complex: certain spins
minimize the single-ion anisotropy energy, while the directions of other spins compromise
between NN, LR, and relativistic anisotropic and
Dzyaloshinsky-Moriya interactions. For the $q_y$$=$$\frac{1}{6}$ structure one can
clearly distinguish different ``domains'' formed by the NN
interactions and the single-ion anisotropy, respectively.
Most importantly, all odd-periodic magnetic structures restore the inversion centers
(associated with the central Mn-sites in Fig.~\ref{fig.SOGS}) and thus exclude any
ferroelectric activity.

\begin{figure}
\begin{center}
\includegraphics[width=3.5cm]{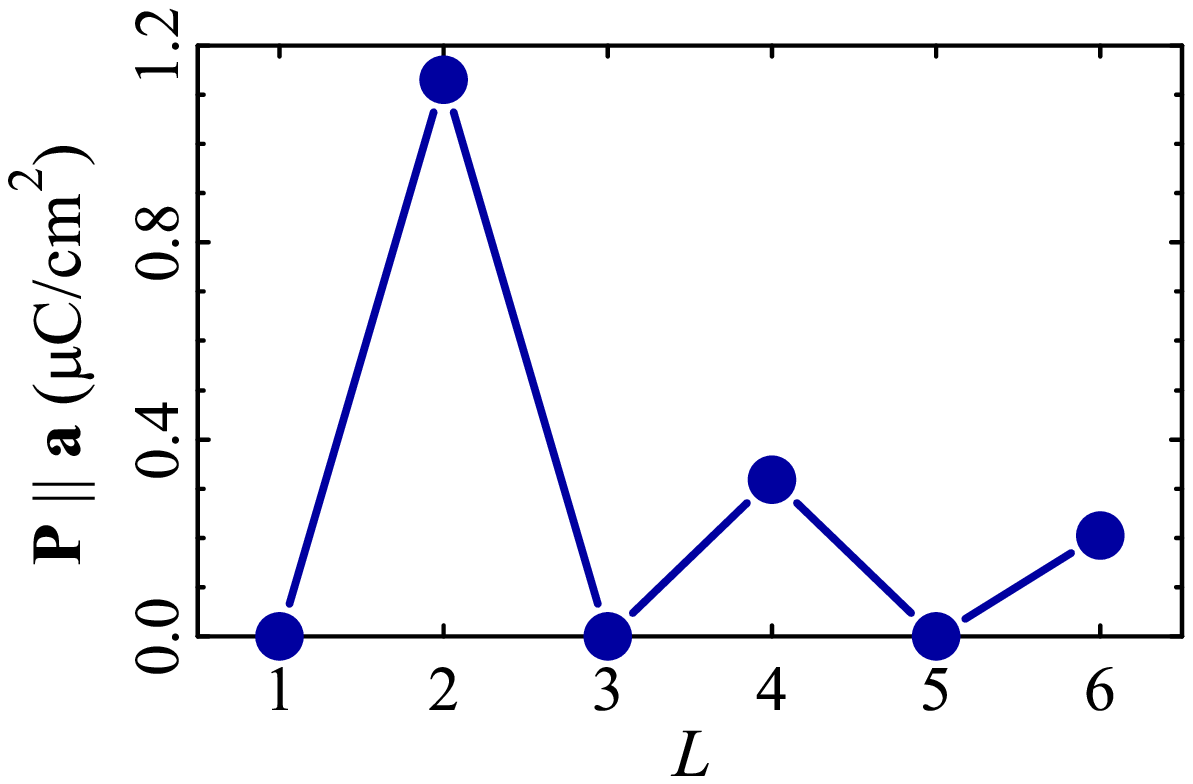}
\includegraphics[width=3.5cm]{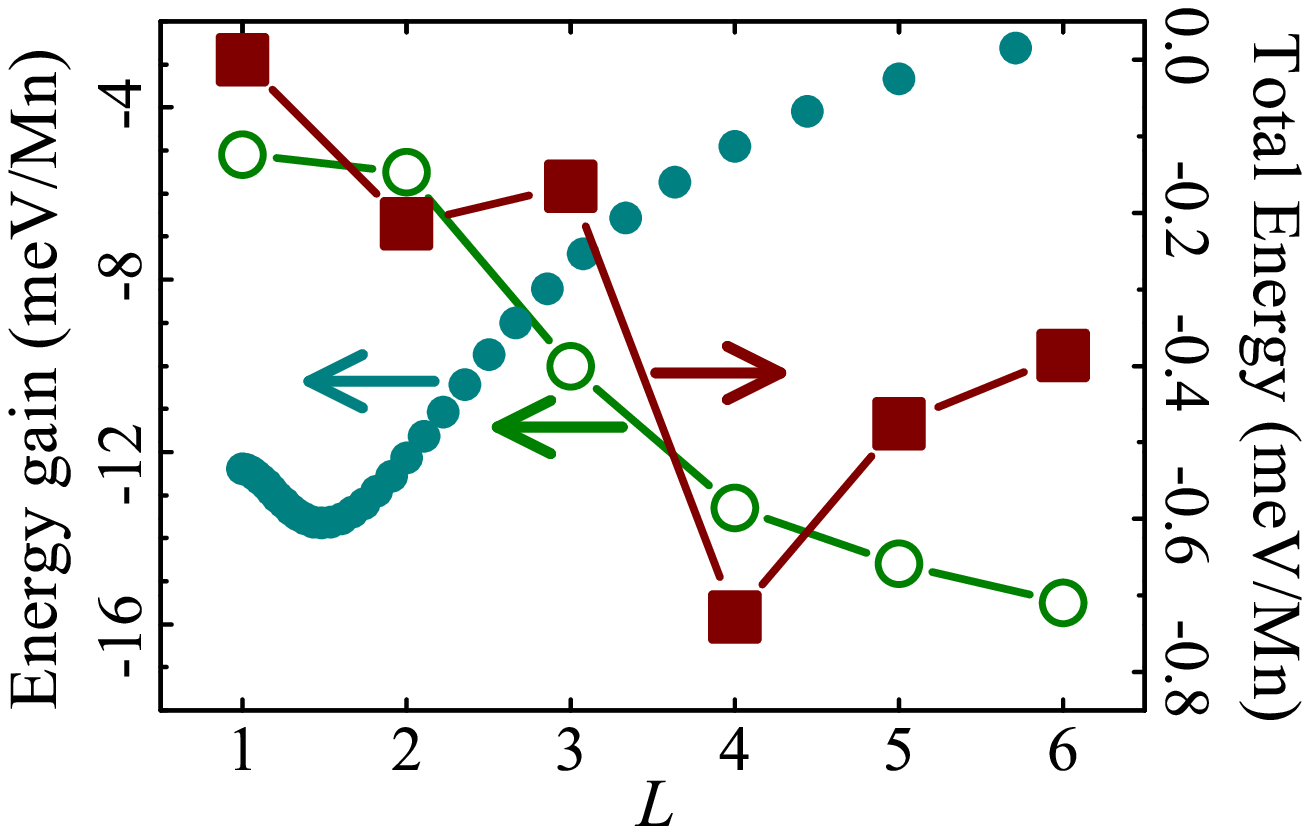}
\end{center}
\caption{\label{fig.TbMnO3PE} Left panel: electronic polarization in TbMnO$_3$
depending on the size of the magnetic unit cell $L$.
Right panel:
dependence of the total energy on $L$$=$$1/q_y$
for the homogeneous spin-spiral states of TbMnO$_3$ without SOI (filled circles),
stabilization energy ($\Delta E$) caused by the
SOI-induced magnetic relaxation (open circles),
and total energies for
magnetic superstructures
shown in Fig.~\protect\ref{fig.SOGS} (squares).}
\end{figure}

  Thus, we predict the oscillatory behavior of ${\bf P}$ depending on
$q_y$$=$$1/L$ (Fig.~\ref{fig.TbMnO3PE}).
The spin-spiral inhomogeneity in the even-periodic structures gives rise to the electronic
polarization, similar to the E-state,
which decreases with the increase of $L$. The canting of spins away from the
collinear arrangement substantially reduces the values of ${\bf P}$.
For example, the electronic polarization
$| {\bf P} |$$=$ 1.1 $\mu$C/cm$^2$, obtained
for the $q_y$$=$$\frac{1}{2}$ structure of
TbMnO$_3$ (Fig.~\ref{fig.TbMnO3PE}),
is reduced by factor two in comparison with the collinear E-state (Fig.~\ref{fig.2b}).
Taking into account the numerical values of $\phi$ (Fig.~\ref{fig.SOGS}), this reduction is
readily explained by the
angle-dependence of ${\bf P}$ in the nonrelativistic case (Fig.~\ref{fig.2b}).
Thus, the SOI only forms the inhomogeneous spin-spiral state, while the
microscopic mechanism yielding finite $ {\bf P} $ is essentially
nonrelativistic.

  The existing theories overestimate the values of
${\bf P}$ (sometimes by an order of magnitude)~\cite{Picozzi,SergienkoPRL},
which is one of the unresolved problems for
the multiferroic manganies. In our calculations
we obtain
${\bf P} || {\bf a}$$=$ 1.4 $\mu$C/cm$^2$ for the
$q_y$$=$$\frac{1}{2}$ structure of HoMnO$_3$ ($\phi$$=$$120^\circ$) and
0.3 $\mu$C/cm$^2$ for the $q_y$$=$$\frac{1}{4}$ structure of
TbMnO$_3$ shown in Fig.~\ref{fig.TbMnO3PE}.
Both values are about three times larger than the
experimental ones~\cite{nonmagneticRE}. Nevertheless,
our theoretical approach also suggests that
it is difficult to obtain a good quantitative agreement because
${\bf P}$ is sensitive to many factors:
(1) the precise value of $\phi$, which itself depends
on the fragile balance of many magnetic interactions;
(2) structural relaxation~\cite{Picozzi,SergienkoPRL};
(3) due to the oscillatory behavior (Fig.~\ref{fig.TbMnO3PE}),
${\bf P}$ can be reduced by
possible
deviations from the even-periodic commensurate alignment.

  Finally, we comment on the relative roles played by the magnetic and structural
relaxation effects in the formation of inhomogeneous magnetic structures.
The behavior of the magnetic stabilization energies $\Delta E$, defined as the energy difference
between fully optimized magnetic structure with SOI and the homogeneous spin-spiral state
without SOI, is explained in Fig.~\ref{fig.TbMnO3PE}.
Since each increase of the supercell provides additional degrees of freedom for the
magnetic relaxation, $\Delta E$ decreases with the increase of $L$.
The absolute value $| \Delta E |$$\sim$6 meV/Mn obtained for the
$q_y$$=$$\frac{1}{2}$ structure of HoMnO$_3$ is comparable
with the energy gain caused by noncentrosymmetric atomic displacements
in the E-state ($\sim$8 meV/Mn~\cite{Picozzi}).
Thus, even for $q_y$$=$$\frac{1}{2}$, the magnetic relaxation cannot be
neglected and should be considered on an equal footing with the structural
relaxation.
Moreover, the transition to the inhomogeneous state is driven by the SOI, prior to
structural relaxation, and in this sense there is no conceptual
difference between HoMnO$_3$ and TbMnO$_3$. Since $| \Delta E |$ increases with $L$, the
relative role of the magnetic relaxation is also expected to increase.
The magnetic relaxation alone readily explains the experimentally observed $q_y$$\approx$$\frac{1}{4}$
periodicity in TbMnO$_3$: $\Delta E$ of the relativistic origin
shifts the total energy minimum of the homogeneous spin-spiral state
($q_y$$=$$0.68$, Fig.~\ref{fig.noSOI}) towards smaller $q_y$.
Thus, the new theoretical minimum corresponds to $q_y$$=$$\frac{1}{4}$ (Fig.~\ref{fig.TbMnO3PE}).
Similar behavior was found for HoMnO$_3$,
which is expected to form similar fourfold periodic
structure. The disagreement
with the experimentally observed twofold periodicity in HoMnO$_3$ is
probably caused by
the neglect of the structural relaxation, which is relatively more important for
$q_y$$=$$\frac{1}{2}$. Quantitative aspects of interplay between
magnetic and structural relaxation effects should be addressed in future theoretical
studies. At present, the structural relaxation cannot be easily implemented
in the present model analysis.

  In summary, we propose that
the homogeneous spin-spiral state in orthorhombic manganites
is deformed by the relativistic SOI. The thus induced inhomogeneity
is responsible for the ferroelectric activity of $R$MnO$_3$.
It would be interesting to check our finding experimentally.

  This work is partly supported by Grant-in-Aid for Scientific
Research (C) No. 20540337 from the
Ministry of Education, Culture, Sport, Science and Technology of
Japan and by Russian Federal Agency for Science and Innovations, grant No. 02.740.11.0217.


\begin{thebibliography}{99}
\bibitem{remark2} The interatomic interactions are defined
as one-half of the HF energy difference between AFM and FM
configurations in each bond.
\bibitem{remark3}
Once the orbital degeneracy is lifted by the JTD,
the HF theory provides a good approximation for the ground-state properties.
The effect of correlation interactions,
which can be treated as a perturbation to the HF ground state~\cite{review2008},
is partially compensated by the magnetic
polarization of the O$2p$ states:
if the former tend to stabilize AFM structures,
the latter favors the FM alignment. Details can be found in Ref.~\cite{JPSJ}.
\end{thebibliography}
\end{document}